\documentstyle[prl,aps,multicol,epsf,preprint]{revtex}

\begin{document} 

\title{
Autocorrelation functions in 3D Fully Frustrated Systems
}  
\author{
G. Franzese$^a$,
A. Fierro$^a$, A. De Candia$^a$
and 
A. Coniglio$^{a,b}$}
\address{
Dipartimento di Scienze Fisiche, Universit\`a di Napoli, 
Mostra d'Oltremare Pad.19 I-80125 Napoli, Italy \\
$^a$INFM - unit\`a di Napoli $~~~$ and $~~~$ $^b$INFN - sezione di Napoli
}

\maketitle 

\begin{abstract} 
We present a numerical study of autocorrelation functions of a 3D Fully 
Frustrated Ising model (FFIM) simulated by 
spin-flip Monte Carlo 
dynamics 
finding simple exponential decay for all the temperature above the critical 
temperature $T_c$ for the autocorrelation of squared magnetization and 
stretched exponential decay for the energy autocorrelation below a temperature 
$T^*$ with $T_c < T^* \leq T_p$ where $T_p$ is the Kasteleyn-Fortuin and 
Coniglio-Klein percolation temperature.
The results are compared to those on 2D FFIM to make light 
on the relevant mechanism in the onset of stretched exponential relaxation 
functions.
\end{abstract} 


Autocorrelation functions of the form
\begin{equation}
f(t)=f_0\exp-(t/\tau)^\beta
\label{eq_stretched}
\end{equation}
called stretched exponential or Kohlrausc-Williams-Watts (KWW) function
are observed in glassy systems and spin glass (SG) well above the respective 
transition temperature and the origin of this kind of behaviour is still matter 
of debate.

For example in SG 
Randeria \emph{ et al.} \cite{ref_randeria} suggested that the onset of 
the non exponential relaxation function coincides with the critical 
temperature of the ferromagnetic Ising model and base their conjecture on 
the presence of non frustrated  ferromagnetic-type clusters of interactions.

Furthermore Huse and Fisher \cite{hf} on the basis of droplet model and 
more recently Stauffer \cite{stauffer} on the basis of percolation arguments
have predicted a KWW autocorrelation function also for ferromagnetic Ising 
model (FIM) in 2D for $T\leq T_c$ where $T_c$ is the critical temperature. 
These arguments show that no such behaviour is expected in more than two 
dimensions and Stauffer has corroborated his prediction with simulations in 2 
and 3D.

It is useful to note at this point that in FIM the critical temperature 
$T_c$ coincides with the 
percolation temperature $T_p$ of the Kasteleyn-Fortuin and Coniglio-Klein 
(KF-CK) clusters. \cite{kf-ck}

These clusters can be obtained introducing bond with probability
$p=1-e^{-2J/(k_BT)}$ between nearest-neighbor pair of spins satisfying the
interaction ($J$ is the interaction strength, $k_B$ the Boltzmann constant
and $T$ the temperature) and represent correlated spins regions.

It would be interesting to investigate whether such prediction also holds for 
frustrated systems where the percolation temperature and the critical 
temperature do not coincide.

In particular in this work we considered the fully frustrated Ising model 
(FFIM). Here we resume the numerical results on 2D \cite{Annalisa} 
and present those in 3D.\cite{tutti}

The FFIM is a model with no disorder (unlike the SG and like the FIM) and with 
the interactions assigned in a regular way to give rise to \emph{frustration}, 
\emph{i.e.} a competition between them (like the SG and unlike the 
FIM).\cite{ff}

The FFIM Hamiltonian is given by
\begin{equation}
H=-J\sum_{\langle{ij}\rangle}(\epsilon_{ij}S_{i}S_{j}-1),
\label{eq_Ising}
\end{equation}
where $\epsilon_{ij}=\pm 1$ are quenched regular assigned variables that give 
rice to frustration  
and $S_i=\pm 1$ are the Ising spins.\cite{Umberto} In this model the KF-CK 
clusters are such that they do not include 
a \emph{ frustrated loop},
that is a closed path of bonds which contains  an odd number of
interactions with $\epsilon_{ij}=-1$. 
These clusters overlap regions of not frustrated interactions and their 
percolation temperature is greater then the critical temperature. In 
Ref.\cite{Umberto} it is studied a general Hamiltonian, with Ising spins and 
$s$-state Potts spins, that has the FFIM as a particular case for $s=1$. It is 
possible to see that for all $s \neq 1$ $T_p$ of KF-CK clusters corresponds to 
a thermodynamic transition of a $s$-state Potts model and that this thermal 
transition disappears for $s=1$, \emph{i.e.} for FFMI. This suggests that the 
KF-CK clusters may have an important role also for FFIM.
 
We simulate this model using 
the standard spin flip \cite{metropolis}
dynamics. 

In the 2D FFIM where $T_p > T_c=0$ recent results \cite{Annalisa} shows 
that stretched exponential autocorrelation functions are observed for 
$T\leq T_p$. Therefore the same qualitatively behaviour of 2D FIM is observed 
but in the FFIM is clear that the onset of stretched exponentials coincides, 
within numerical precision, with $T_p\neq T_c$, while in FIM is $T_p=T_c$. 

To put more light on the problem we have done simulations on 3D FFIM. 

To this goal we have calculated the autocorrelation function of squared 
magnetization $M^2$ and of energy $E$, where 
for a generic observable $A$ the relaxation function is defined as
\begin{equation}
f_A(t)=\frac{\langle\delta A(t)\delta A(0)\rangle}
{\langle(\delta A)^2\rangle},
\end{equation}
where $\delta A(t)= A(t)-\langle A\rangle$, $t$ is the time and 
$\langle \cdot \rangle$ is the thermal average. 

We fit our data with a general function
\begin{equation}
f(t)=f_0\frac{\exp-(t/\tau)^\beta}{t^x}
\label{eq_str_pl}
\end{equation}
where $f_0$, $\tau$, $\beta$ and $x$ are parameters.

\begin{center}
\begin{figure}
{\Huge a} \mbox{ \epsfxsize=6.3cm \epsfysize=11cm \epsffile{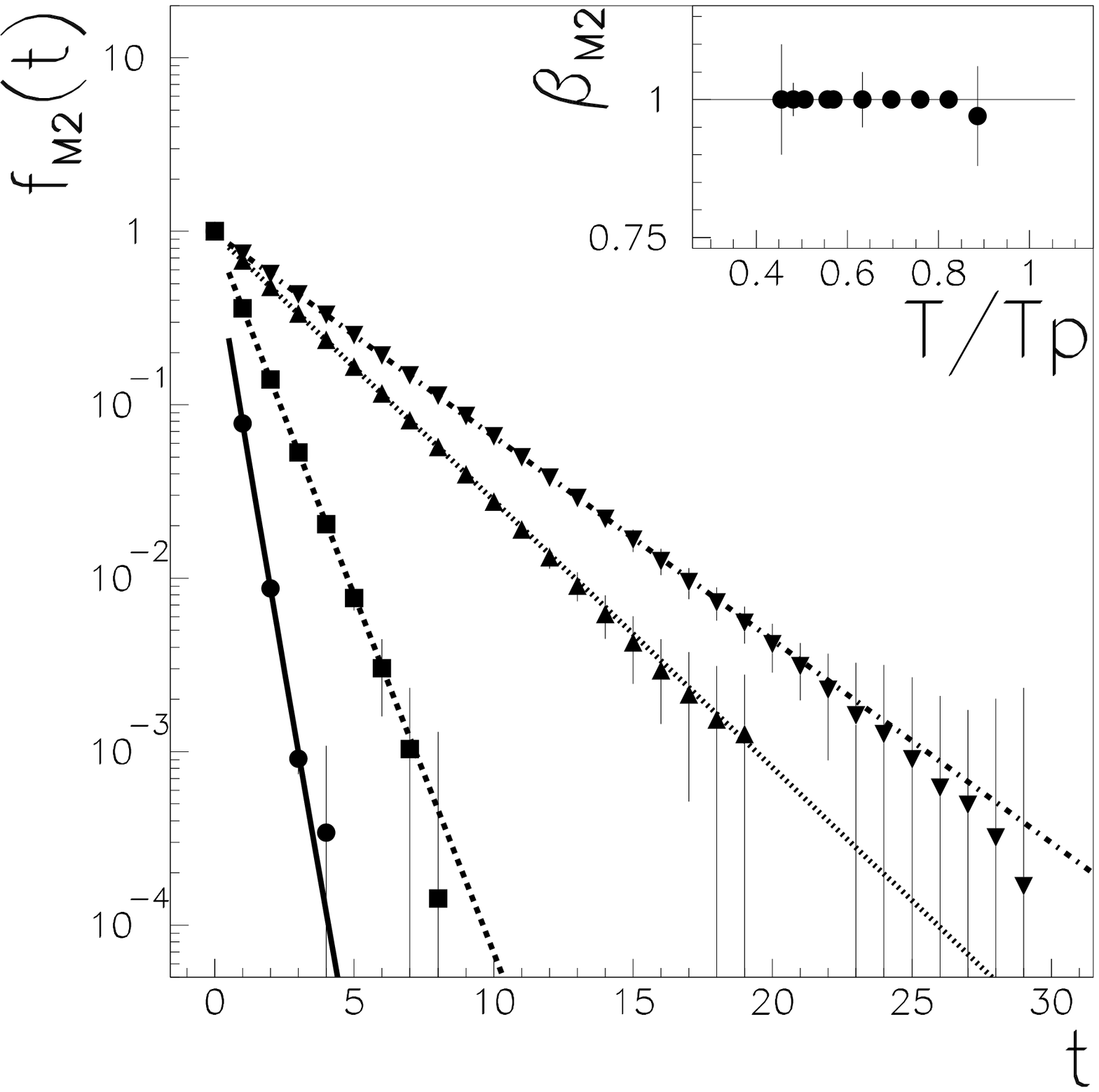} }
{\Huge b} \mbox{ \epsfxsize=6.3cm \epsfysize=11cm \epsffile{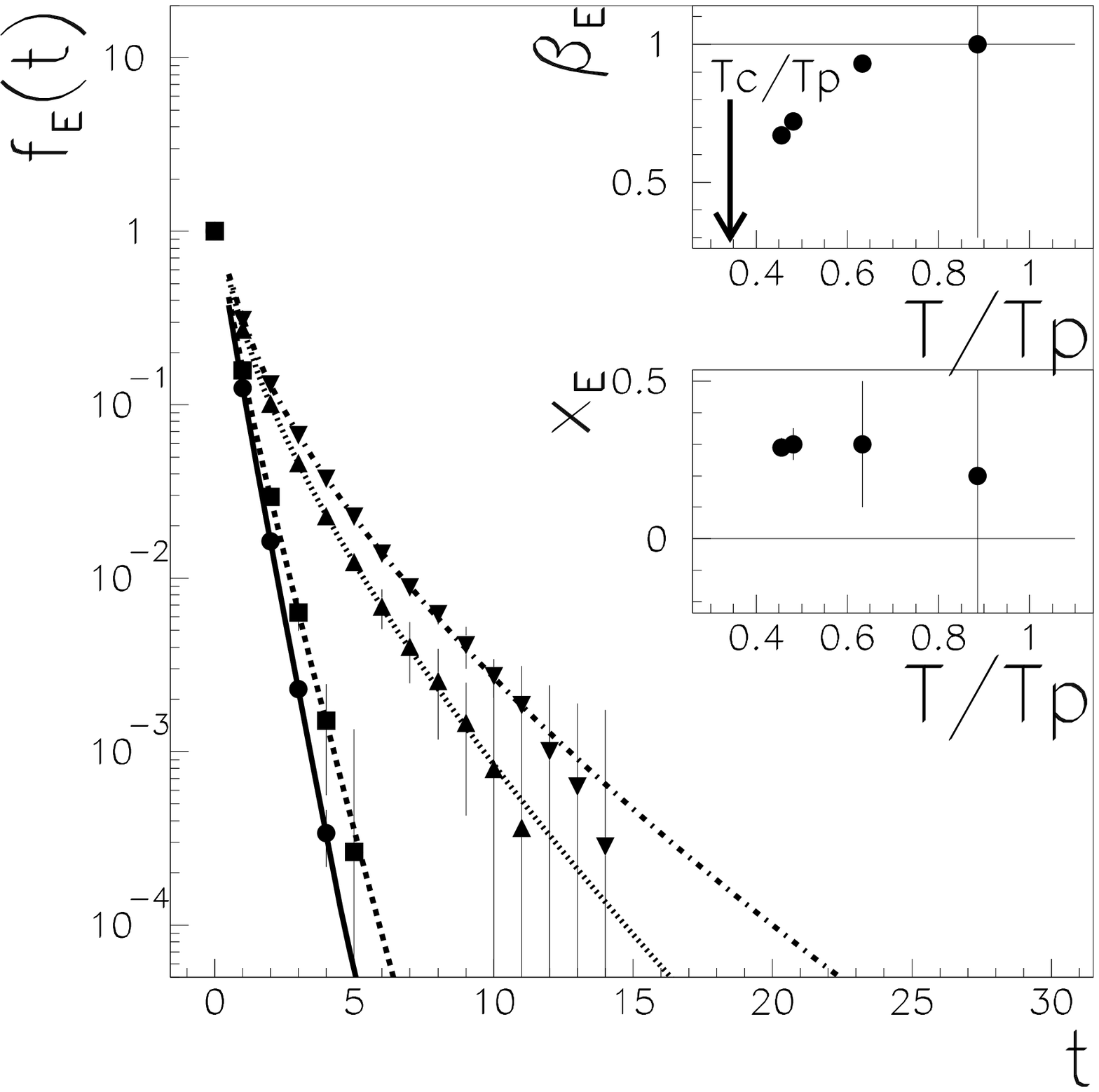} } 
\caption{
Correlation functions for $L=20$:
(a) $f_{M^2}$ and (b) $f_E$ at (from left to right) 
$k_BT/J=3.5$, 2.5, 1.9, 1.8 from respectively 2,7,8,7 independent 
runs with statistics of $5 \cdot 10^6$ $L^3$ 
equilibrium update after $5 \cdot 10^3$ $L^3$ update of thermalization;
insets: the corresponding estimated $\beta$ and $x$ in eq.(4) vs.
$T/T_p(L)$ where $k_BT_p(L=20)/J\simeq 3.95$.
[10]
}
\label{results} 
\end{figure}
\end{center}

In Fig.\ref{results} we summarize the results 
for systems with linear size $L=20$ and for temperatures below the 
estimated \cite{Giancarlo} $T_p(L)$ of KF-CK clusters in cubic FFIM.

For $f_{M^2}$ the estimated parameters are $\beta\simeq 1$ and $x\simeq 0$ 
for all simulated temperatures below $T_p(L)$. 
For $f_E$ there is a temperature $T^*$, between $T_p(L)$ and 
$T_c\simeq 1.355 J/k_B$ (the estimated critical temperature \cite{Diep}), 
below which is $\beta < 1$ and $x$ weakly changes. Simulations 
on $L=30 \div 50$ confirm these data.

Therefore
the 3D FFIM results
show a simple exponential decay in $M^2$ correlation functions 
but a stretched exponential decay in $E$ correlation functions for $T\leq T^*$
with $T_c< T^* \leq T_p$. In conclusion the results on $f_E$ in 3D FFIM, even if
less clear than those on 2D FFIM, confirm the existence of an onset $T^*$ of 
stretched exponential for the relaxation function well above the critical 
temperature $T_c$ and almost equal (or slightly below) to the KF-CK percolation
temperature $T_p$ that, in a generalized model with also $s$-state Potts
variables with $s\neq 1$, corresponds to a thermodynamic transition of the 
Potts variables.\cite{Umberto}

\end{document}